\documentclass[epjST]{svjour}

\usepackage{graphics,amsmath}

\usepackage[sort,compress]{cite}

\newcommand{\ol}[1]{\overline{#1}}

\newcommand{\F}{\mathbf{F}}
\newcommand{\G}{\mathbf{G}}
\renewcommand{\H}{\mathbf{H}}

\newcommand{\rbf}{\mathbf{r}}
\newcommand{\tbf}{\mathbf{t}}
\newcommand{\vbf}{\mathbf{v}}
\newcommand{\x}{\mathbf{x}}

\begin{document}
\title{Hydrodynamic synchronization of flagellar oscillators}
\author{
B.M. Friedrich\inst{1,2}\fnmsep
\thanks{\email{benjamin.m.friedrich@tu-dresden.de}}
}
\institute{
(1) Max Planck Institute for the Physics of Complex Systems, Dresden, Germany\\
(2) TU Dresden, cfaed, Dresden, Germany}

\abstract{
In this review, we highlight the physics of synchronization in collections of beating cilia and flagella.
We survey the theory synchronization in collections of noisy oscillators. 
This framework is applied to flagellar synchronization by hydrodynamic interactions. 
The time-reversibility of hydrodynamics at low Reynolds numbers prompts swimming strokes that break symmetry 
to facilitate hydrodynamic synchronization. 
We discuss different physical mechanisms for flagellar synchronization, 
which break this symmetry in different ways. 
} 

\maketitle

\section{Introduction}
\label{intro}

What we call today synchronization of oscillators
was first described in 1665 by van Huygens,
when he observed two pendulum clocks adopting a common rhythm that lasted for many hours \cite{Pikovsky:synchronization}. 
This was surprising, given the notorious inaccuracy of clocks at the time. 
In fact, a weak mechanical coupling had entrained the two pendulum clocks.
A simple class-room demonstration of this experiment can be realized using 
two metronomes on a swinging tray \cite{Pantaleone:2002}.
In the middle of the $20^\mathrm{th}$ century, work on 
phase-locked electronic oscillators motivated the development 
of a general theory of synchronization \cite{Adler:1946,Stratonovich:1963}, 
which we briefly review in section 2.

Synchronization of active oscillators has since then been observed in a number of examples, 
including synchronization of biological oscillators.
As a popular example, synchronization of walking gaits 
among pedestrians was observed on the Millennium bridge in central London, 
which caused large-amplitude vibrations of the bridge \cite{Dallard:2001,Strogatz:2005}.
This synchronization was the result of a mechanical coupling. 
Walking of individual pedestrians excited small-amplitude vibrations;
in turn, each pedestrian adapted its frequency and phase of walking to the vibrations of the bridge.

Synchronization of biological oscillators is also observed at the cellular scale.
Many cells are propelled in a liquid by 
regular bending waves of slender cell appendages termed cilia and flagella \cite{Brennen:1977,Lauga:2009,Elgeti:2015}.
The rhythmic beat of a cilium or flagellum represents a prime example of a biological oscillator. 
Cilia and flagella share a highly conserved core, the axoneme, which comprises a 
regular cylindrical arrangement of 9 doublet microtubules, which are connected by $10^4-10^5$ dynein molecular motors \cite{Alberts:cell,Nicastro:2006}.
These molecular motors convert chemical energy in the form of ATP into mechanical work to bend the axoneme \cite{Brokaw:1989}.
A dynamic instability of collective motor dynamics gives rise to self-organized bending waves of cilia and flagella \cite{Lindemann:1994,Riedel:2007,Brokaw:2008}. 
The distinction between eukaryotic cilia and eukaryotic flagella is mainly made with respect to their length and function,
and we will use the same term flagellum for both. 
Note, however, that the eukaryotic flagellum is very different from the bacterial flagellum, which is a passive protein filament. 

Collections of flagella can adopt a common rhythm and synchronize their beat.
It had been observed already by Gray almost 100 years ago that pairs of sperm cells swimming in close proximity can synchronize their flagellar beat 
\cite{Gray:1928,Woolley:2009}.
At high sperm densities, swimming sperm can self-organize into vortex patterns with vortices of about 5 cells 
with mutually phase-locked flagellar beat \cite{Riedel:2005a}.
Flagellar synchronization plays an important role for the collective dynamics in carpets of short flagella termed cilia, 
which can phase-lock their beat to give rise to propagating metachronal waves.
This coordinated flagellar beating results in effective surface slip and thereby facilitates self-propulsion 
of multi-ciliated organisms such as the unicellular eukaryote \textit{Paramecium} \cite{Machemer:1972}
or the green alga colony \textit{Volvox} \cite{Brumley:2012}.
Inside the mammalian body, dense carpets of cilia on epithelial surfaces pump fluids, 
such as mucus in mammalian airways or cerebrospinal fluid in the brain \cite{Sanderson:1981}.
Again, these cilia synchronize their beat to exhibit metachronal waves.
This self-organized ciliar dynamics is important for efficient fluid pumping \cite{Cartwright:2004,Osterman:2011,Elgeti:2013}. 

The bi-flagellated, unicellular green alga \textit{Chlamydomonas} 
is serving as a model organism to study physical mechanisms of flagellar synchronization \cite{Polin_EPJST}, 
which is singled out by the ease of its experimental handling and accessibility for quantitative study 
\cite{Ruffer:1998a,Polin:2009,Goldstein:2009,Goldstein:2011,Geyer:2013}.
\textit{Chlamydomonas} swims like a breast-stroke swimmer with two flagella, 
which can synchronize their beat,
both in free-swimming cells and cells held in a micro-pipette.
On physical grounds, this flagellar coordination is surprising, 
since there is no direct coupling between the respective molecular motors that drive the beat of each flagellum.
There is also no evidence for a chemical master oscillator that could set a common rhythm for both flagella.
Below, we discuss different physical mechanisms that can 
account for the stabilization of synchronous flagellar beating by a mechanical coupling.
In a generic description of synchronization in pairs of oscillators known as the Adler equation discussed below, 
there are two possible synchronization states: in-phase and anti-phase synchronization. 
In \textit{Chlamydomonas}, in-phase synchronization corresponds to a mirror-symmetric, breaststroke-like dynamics of the two flagella
and anti-phase synchronization to a phase-shift of approximately $\pi$ between both flagella.
While wild-type \textit{Chlamydomonas} cells usually display in-phase flagellar synchrony, 
stable anti-phase synchronization has been observed in a flagellar mutant \cite{Leptos:2013}.
For \textit{Chlamydomonas}, 
synchronized beating is required to swim both fast and straight \cite{Polin:2009,Geyer:2013}.

The flagellar beat responds to external mechanical forces, such as hydrodynamic friction forces. 
It was observed that the beat of sperm cells slowed down 
upon an increase in the viscosity of the swimming medium \cite{Brokaw:1975,Friedrich:2010}.
Friedrich \textit{et al.} measured a dynamic force-velocity relationship of the flagellar beat
for \textit{Chlamydomonas} cells by correlating changes of the flagellar phase speed during a beat cycle
with a rotational motion of these cells, which imparted known hydrodynamic forces on each flagellum \cite{Geyer:2013}.
Transient external flows were shown to reversibly perturb a limit cycle of flagellar bending waves \cite{Wan:2014}.
Using strong external flows it was even possible to stop and re-start flagellar oscillations in a reversible manner \cite{Klindt:arxiv}.
This susceptibility of the flagellar beat to external forces allows for the synchronization of flagellar beating to 
an oscillatory driving.
Flagellar beating could be entrained to periodic mechanical forcing using vibrating micro-needles \cite{Okuno:1976} 
or oscillatory external flows \cite{Quaranta:2015}.
It has been proposed already by Taylor that a weak mechanical coupling between flagella 
could also underly the striking phenomenon of spontaneous synchronization in collections of cilia and flagella.

Taylor considered direct hydrodynamic interactions between flagella
as one possibility for a mechanical coupling between beating flagella \cite{Taylor:1951}.
Direct hydrodynamic interactions between flagella refer to hydrodynamic flows generated by one flagellum
that propagate to a nearby flagellum and exert hydrodynamic friction forces on it.
This concept of hydrodynamic synchronization sparked a continuous research effort, both experimental and theoretical.
A clear demonstration of synchronization by direct hydrodynamic interactions 
was achieved only recently in an experimental setup of two flagellated cells held by separate micro-pipettes at a distance \cite{Brumley:2014}.
In addition to synchronization by direction hydrodynamic interactions, 
a mechanism of mechanical self-stabilization was proposed for free-swimming cells \cite{Friedrich:2012c,Geyer:2013,Bennett:2013}.

Important insights into physical mechanisms of 
hydrodynamic synchronization have been gained by the study of artificial micro-swimmers and micro-actuators, 
such as colloids driven by oscillating magnetic fields or 
`light-mill' micro-rotors driven by laser light, 
which likewise
perform cyclic motions  
\cite{Kotar:2010,Bruot:2011,Bruot:2012,Leonardo:2012,Lhermerout:2012}. 
Several of these oscillators can synchronize their oscillations
through a weak coupling mediated by hydrodynamic interactions.

Synchronization by hydrodynamic interactions is a subtle phenomenon that requires broken symmetries 
\cite{Vilfan:2006,Niedermayer:2008,Elfring:2009,Polotzek:2013,Elgeti:2015}. 
The physical reason for this is the time-reversal symmetry of the Stokes equation, 
which governs hydrodynamics in the low Reynolds number regime of flagellar flows \cite{Lauga:2009}. 
As a consequence, active oscillators such as beating flagella must break
either spatial or temporal symmetries for synchronization to occur.
We will discuss different physical mechanisms for flagellar synchronization \cite{Vilfan:2006,Niedermayer:2008,Uchida:2011,Theers:2013}, 
which break this symmetry in different ways. 

In section 2 of this review, 
we will discuss the nonlinear dynamics of synchronization of noisy oscillators.
In section 3,
we will apply this theory in the context of flagellar synchronization, 
and discuss the role of broken symmetries for synchronization.

\section{Nonlinear dynamics of synchronization}

\subsection{Active oscillators}

We consider a prototype of an active oscillator, the normal form of a Hopf oscillator 
with complex oscillator variable $Z=A\exp(i\varphi)$ \cite{Crawford:1991}
\begin{equation}
\label{eq_hopf}
\dot{Z} = i \omega Z + \mu (A_0^2 - |Z|^2) Z.
\end{equation}
The onset of flagellar oscillations 
had been previously described as a supercritical Hopf bifurcation \cite{Camalet:2000}
with normal form given by Equation (\ref{eq_hopf}).
Note that the nonlinear friction term $\mu(A_0^2-|Z|^2)$ becomes negative for $|Z|<A_0$.
Negative friction is a common feature of active systems characterized by positive feedback loops.
In these systems, a reservoir of energy becomes depleted to drive the oscillations, 
with energy eventually dissipated as heat. 

At steady-state, Equation (\ref{eq_hopf}) implies sustained oscillations $Z=A_0\exp(i\varphi)$ with amplitude $A_0$.
The dynamics of the phase $\varphi$ is given by 
\begin{equation}
\label{eq_phi}
\dot{\varphi} = \omega .
\end{equation}
This simple description of a phase oscillator generalizes to a large class of nonlinear oscillators,
and proves to be particularly useful if the amplitude of oscillations stays approximately constant. 
It should be emphasized that $\varphi$ is only defined up to multiples of $2\pi$ and that 
any physical quantity that depends on $\varphi$ must in fact be a $2\pi$-periodic function. 
In real applications, one would compute $Z$ and its phase $\varphi$ using 
either the Hilbert transform of $x$ (analytical signal), 
or by limit-cycle reconstruction as discussed now. 

\begin{figure}
\begin{center}
\resizebox{0.75\columnwidth}{!}{%
\includegraphics{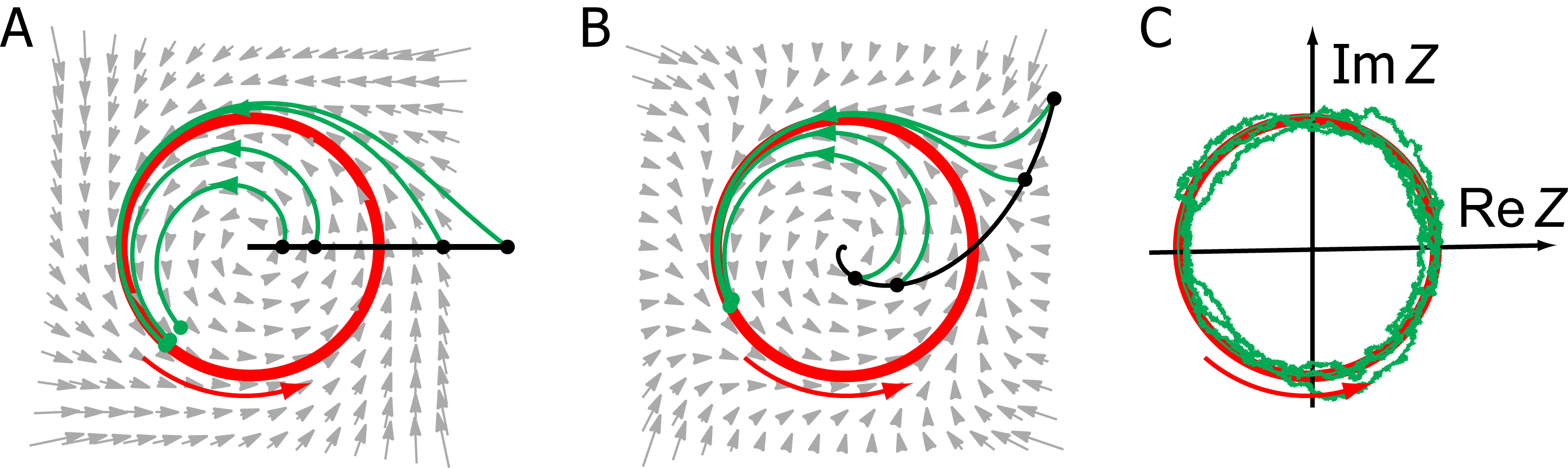}
}
\end{center}
\caption[]{
\textit{Phase-space of a limit cycle oscillator.}
\textbf{A.} 
The red trajectory represents a limit cycle, 
\textit{i.e.} a stable, periodic orbit of the dynamics.
Example trajectories (green) that start in the vicinity of this limit cycle will eventually converge onto this limit cycle.
The black line indicates a so-called isochrone: 
trajectories that start on the same isochrone will converge on the same point on the limit cycle. 
The flow-field is indicated by gray arrows.
\textbf{B.} 
Same as panel A, but for a non-isochronous limit cycle oscillator
with non-radial isochrones.
The simulated dynamics corresponds to a non-isochronous Hopf oscillator with
$\dot{Z} = i \omega Z + (\mu + i\omega_1) (A_0^2 - |Z|^2) Z$, 
where the isynchrony parameter $\omega_1=0$ in panel A and $\omega_1>0$ in panel B.
\textbf{C.} 
Noisy limit cycle oscillations
of a complex oscillator variable $Z$, see also Equation (\ref{eq_hopf_noise}).
}
\label{figure_limit_cycle}
\end{figure}

We define an oscillator as a stable limit cycle, 
\textit{i.e.} a periodic orbit such that $\x(t+T)=\x(t)$ for some period $T$, 
for which all trajectories 
starting in a sufficiently thin tubular neighborhood of the orbit will eventually 
converge to this orbit.
This is illustrated in Figure \ref{figure_limit_cycle}. 
Note that every limit cycle can be parameterized by a phase variable $\varphi$,
such that $\varphi$ advances by $2\pi$ during one cycle. 
We require that the phase variable advances with uniform speed along the limit cycle. 
Given an arbitrary parametrization of a limit cycle by a cyclic variable $\theta$, 
this can always be achieved by an appropriate reparametrization $\varphi=\varphi(\theta)$ \cite{Kralemann:2008,Schwabedal:2013}. 
The phase parameterization is unique up to the choice of start point.
The phase parameterization can be extended to a tubular neighborhood of the limit cycle (by defining so-called iso-chrones) \cite{Pikovsky:synchronization}. 
Thereby, the nonlinear dynamics in the vicinity of a limit cycle 
can be mapped on the simple phase oscillator description given in Equation~(\ref{eq_phi}).

This definition generalizes in a straight-forward manner to stochastic dynamical systems, 
where the presence of noise implies that trajectories fluctuate around the limit cycle. 
The variance of amplitude fluctuations is set by a competition of 
the `attractor strength' of the limit cycle (Lyapunov exponent) and the noise strength. 
An example of such noisy oscillations is given by a (special case) of 
a noisy Hopf oscillator with complex oscillator variable $Z$ \cite{Stratonovich:1963,Ma:2014}
\begin{equation}
\label{eq_hopf_noise}
\dot{Z} = i \omega Z + \mu (A_0^2 - |Z|^2) Z + Z(\xi_A+i\xi_\varphi),
\end{equation}
where $\xi_A$ and $\xi_\varphi$ are uncorrelated Gaussian white noise variables 
satisfying 
$\langle \xi_i(t)\xi_j(t')\rangle = 2D_j \delta_{jk} \delta(t-t')$, 
$i,j\in\{A,\varphi\}$,
with phase and amplitude noise strength $D_\varphi$ and $D_A$. 
Here, Stratonovich calculus is used.
In the limit of weak noise, we find effective Equations for amplitude and phase. 
Amplitude fluctuations are governed by a noisy relaxation process
that always reverts to the mean amplitude $A_0$ \cite{Risken}.
In the limit of weak noise, we obtain an Ornstein-Uhlenbeck process
reverting to the mean amplitude $A_0$, 
\begin{equation}
\dot{A} = 2\mu A_0^2(A_0-A) + A_0\xi_A, 
\end{equation}
while the noisy phase dynamics is characterized by phase diffusion with 
phase diffusion coefficient $D=D_\varphi$
\begin{equation}
\label{eq_phi_noise}
\dot{\varphi} = \omega+\xi_\varphi.
\end{equation}
For simplicity, 
Equation (\ref{eq_hopf_noise}) refers to an isochronous oscillator with isochrony parameter $\omega_1=0$.
In the general case of a non-isochronous oscillator, 
the coefficient $\mu$ has to be replaced by $\mu+i\omega_1$.
In this case, an effective phase diffusion coefficient
$D=D_\varphi+(\omega_1 A_0/\mu)^2 D_A$ is found
which combines the contribution from phase fluctuations and amplitude fluctuations.

In applications, the phase diffusion coefficient $D$ can be inferred from the width $\Delta\omega=2D$ at half-maximum 
of the Fourier peak of the power spectral density of $Z$, 
or from the inverse exponential decay time of the phase correlation function
$|\langle \exp[i\varphi(t+\Delta t)-i\varphi(t)]\rangle|=\exp(-D\, \Delta t)$.

\subsection{The Adler Equation for a pair of coupled oscillators}

\begin{figure}
\begin{center}
\resizebox{0.75\columnwidth}{!}{%
\includegraphics{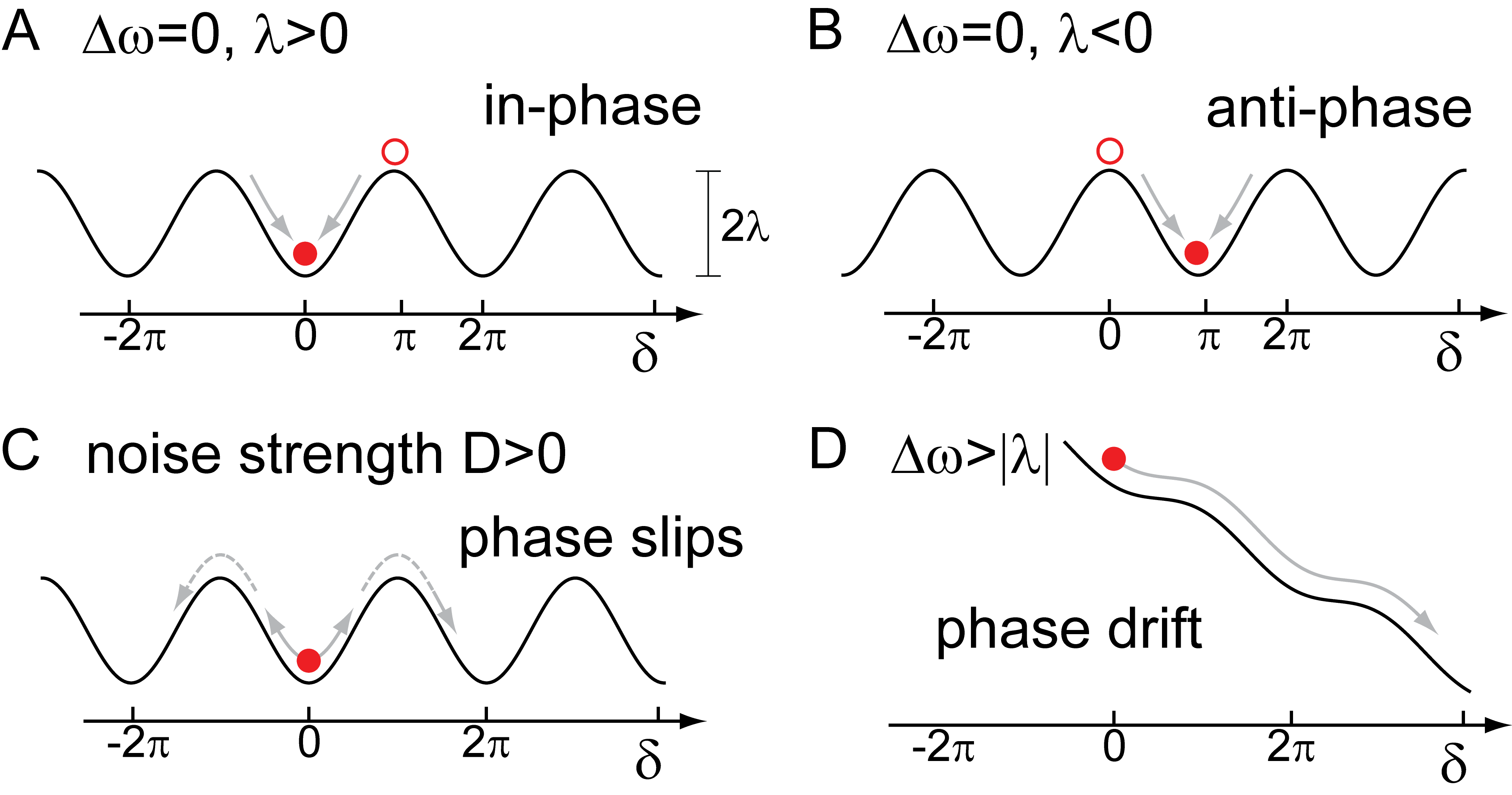}
}
\end{center}
\caption[]{
\textit{Dynamic regimes of the Adler equation.}
The Adler equation
for the phase difference $\delta$ between two coupled oscillators 
can be interpreted as the dynamics of an over-damped particle moving in a tilted periodic potential 
$U(\delta)=-\Delta\omega\delta-\lambda\cos\delta$ \cite{Goldstein:2009}.
\textbf{A.}
For positive synchronization strength $\lambda>0$ and zero frequency mismatch $\Delta\omega=0$, 
in-phase synchronization with $\delta=0$ is stable, 
while anti-phase synchronization with $\delta=\pi$ is unstable.
\textbf{B.}
For negative synchronization strength $\lambda<0$, the stability behavior is reversed:
in-phase synchronization is unstable, while anti-phase synchronization is stable. 
\textbf{C.}
In the presence of noise, the phase difference $\delta$ will fluctuate around its mean position. 
Occasionally, phase slips occur during which the phase difference changes by $\pm 2\pi$.
\textbf{D.}
In case the frequency mismatch exceeds the synchronization strength, $\Delta\omega>|\lambda|$, 
no synchronized states can exist, and the system displays phase drift with a phase difference that 
increases (decreases) monotonically.
}
\label{figure_adler}
\end{figure}

We consider two phase oscillators, 
which are coupled by a generic coupling term $C(\delta)$
\begin{eqnarray}
\label{eq_two_oscillators}
\dot{\varphi}_L &= \omega_L + C(\varphi_L-\varphi_R), \\
\dot{\varphi}_R &= \omega_R + C(\varphi_R-\varphi_L).
\end{eqnarray}
For mechanical oscillators, such a coupling could arise for example 
by elastic or hydrodynamic interactions. 
We are interested in the dynamics of the phase-difference 
$\delta=\varphi_L-\varphi_R$. 
To be physically sound, the coupling term $C(\delta)$ 
must be a $2\pi$-periodic function of $\delta$.
Thus, we can expand $C(\delta)$ as a Fourier series, 
$C(\delta)=\sum_n C'_n \cos(n\delta) + C''_n \sin(n\delta)$.
We find that the dynamics of $\delta$ depends only on the odd coupling terms, 
$\dot{\delta} = \Delta\omega + 2 \sum_n C''_n \sin(n\delta)$, 
where $\Delta\omega=\omega_L-\omega_R$.
In many applications, $C(\delta)$ is dominated by its principal Fourier mode, 
allowing us to write
\begin{equation}
\dot{\delta} = \Delta\omega - \lambda \sin(\delta).
\label{eq_Adler}
\end{equation}
Equation (\ref{eq_Adler}) is the well-known Adler equation \cite{Adler:1946}, 
where we have set $\lambda=-2C''_1$.
In the special case of zero frequency mismatch, $\Delta\omega=0$, 
we can readily read off the steady states 
of the Adler Equation (\ref{eq_Adler}) as
$\delta=0$ and $\delta=\pi$.
In-phase synchronization with $\delta=0$ is stable for $\lambda>0$ and
unstable for $\lambda<0$, see also Figure \ref{figure_adler}.
A small frequency mismatch $|\Delta\omega|>0$ between both oscillators will detune 
the steady state phase difference to a value
$\delta=\sin^{-1}(\Delta\omega/\lambda)$, provided $|\Delta\omega|<\lambda$.
If the phase-difference becomes too large, $|\Delta\omega|>\lambda$,
Equation (\ref{eq_Adler}) does not possess a steady state anymore. 
This implies that synchronization cannot occur, 
corresponding to a case of phase drift, where $\delta$ increases without bound.

We can account for synchronization in the presence of noise by considering
the stochastic Adler equation \cite{Adler:1946,Stratonovich:1963,Pikovsky:synchronization}
\begin{equation}
\dot{\delta} = \Delta\omega - \lambda \sin(\delta)+\xi.
\label{eq_Adler_stoch}
\end{equation}
Here, $\xi(t)$ is a Gaussian white noise variable satisfying
$\langle \xi(t)\xi(t') \rangle = 2D\, \delta(t-t')$, 
where $D=D_1+D_2$ denotes the sum of the phase-diffusion coefficients of the individual phase oscillators $\varphi_L$ and $\varphi_R$.
In the presence of noise, the phase-difference $\delta$ has
stationary probability distribution $p(\delta)$ of finite width;
for zero frequency mismatch, $\Delta\omega=0$, this distribution reads \cite{Stratonovich:1963}
\begin{equation}
p(\delta) = \frac{1}{2\pi I_0(\lambda / D)} \exp\left( \frac{\lambda}{D}\cos\delta \right),
\end{equation}
where $I_0$ in the normalization factor denotes a modified Bessel function of the first kind.
Noise induces occasional phase-slips, during which one of the two oscillators will perform an extra cycle. 
The presence of a frequency mismatch introduces a bias in the rates $G_{+}$ and $G_{-}$ of phase-slips 
in the directions 
$\delta\rightarrow\delta+2\pi$ and
$\delta\rightarrow\delta-2\pi$, respectively
\cite{Stratonovich:1963}
\begin{equation}
\label{eq_G} 
G_{\pm} = \frac{D}{4\pi^2}\, |I_{i\Delta\omega/D}(\lambda/D)|^{-2} \exp\left( \pm \frac{\Delta\omega}{D} \right).
\end{equation}
Here, $I_{ix}$ denotes the modified Bessel function of the first kind with imaginary index.
Equation (\ref{eq_G}) implies that the rate of phase-slips is exponentially suppressed for $D\ll |\lambda|$
while for $D\gg|\lambda|$ the dynamics of $\delta$ is virtually indistinguishable from phase diffusion. 
For $\Delta\omega=0$, we note the asymptotic formulas
\begin{equation}
G_+=G_-=
\begin{cases}
\exp(-2\lambda/D)/(2\pi) & D\ll |\lambda| \\
D/(2\pi)^2 & D\gg|\lambda|
\end{cases}.
\end{equation}
Many additional analytical results are known, see \textit{e.g.} the book by Stratonovich \cite{Stratonovich:1963}. 
% Stratonovich: (9.94), p. 255
% A0=1, K=2D_phi, \Delta=\Delta\omega, \Delta_s=\lambda, D=2A0^2\Delta/K=\Delta\omega/D_phi, Ds=2A0^2\Delta_s/K=\lambda/D_phi

In conclusion, a generic anti-symmetric coupling between two oscillators implies their mutual synchronization, 
provided this coupling is sufficiently strong to overcome both a possible frequency mismatch and the effect of noise.
In the minimal model given by Equation (\ref{eq_Adler}), 
there exists an in-phase and an anti-phase synchronized state.
Which of the two is stable (and thus will be observed in experiments) depends on the sign 
of the synchronization parameter $\lambda$
and is thus a non-generic system's property that depends on the particularities of the given system. 
    
\subsection{The Kuramoto model of $N$ coupled oscillators}

We now consider the general case of $N$ oscillators with mutual coupling. 
In this case, the Adler Equation (\ref{eq_Adler}) generalizes to the famous Kuramoto model \cite{Kuramoto:1984}
\begin{equation}
\dot{\varphi}_i = \omega_i - \frac{\lambda}{N} \sum_{k=1}^N \sin(\varphi_i-\varphi_j),
\label{eq_kuramoto}
\end{equation}
with coupling parameter $\lambda>0$.
Here, we assumed an all-to-all coupling of the oscillators.
The Kuramoto model provides a minimal description of oscillator arrays with long-range coupling
that is analytically tractable.

We assign a complex oscillator variable $Z_k=\exp(i\varphi_k)$ to each phase oscillator $\varphi_k$. 
The mean of all $Z_k$ specifies an order parameter $r=|\ol{Z}|$ of phase coherence
\begin{equation}
\ol{Z} = \frac{1}{N} \sum_{k=1}^N Z_k = r\,\exp(i\psi) .
\label{eq_r}
\end{equation} 
With this definition, Equation (\ref{eq_kuramoto}) becomes
\begin{equation}
\dot{\varphi}_i = \omega_i - \lambda r \sin(\psi-\varphi_i),
\label{eq_kuramoto_mean_field} 
\end{equation} 
\textit{i.e.} we can replace the coupling between the individual oscillators
by a coupling to a mean-field.
Many analytical results are known, 
including a complete set of integrals of motion for the special case of identical oscillator frequencies \cite{Watanabe:1993}.

We consider the thermodynamic limit of $N\rightarrow\infty$, 
with a given distribution of oscillator frequencies $p(\omega)$.
Any oscillator $Z$ for which 
$|\Delta\omega|<\lambda r$
will phase-lock to $\ol{Z}$ 
with a constant phase-lag $\Delta\varphi=\sin^{-1}[\Delta\omega/(\lambda r)]$.
Here, $\Delta\omega=\omega-\omega_0$ 
denotes the mismatch between the intrinsic oscillator frequency $\omega$ and 
the global synchronization frequency $\omega_0=\dot{\psi}$.
Thus, the stationary probability density
of the oscillator phase $\varphi$, conditioned by the global phase $\psi$,  
simply reads $\rho_{s}(\varphi|\psi)=\delta(\varphi-\psi-\Delta\varphi)$ in this case.
Those oscillators with $|\Delta\omega|>\lambda r$ will not phase-lock to $\ol{Z}$ and display phase drift instead.
From Equation (\ref{eq_kuramoto_mean_field}), 
one finds for the stationary probability density
of their phase 
$\rho_{u}(\varphi|\psi) \sim \dot{\varphi}-\omega_0 = \Delta\omega + \lambda r\sin(\Delta\varphi)$.
By inserting the combined stationary probability density $\rho(\varphi|\psi)$ into Equation (\ref{eq_r}),
one obtains an implicit equation for the order parameter $r$ that depends on $p(\omega)$.
For the special case of a Lorentzian distribution of oscillator frequencies 
with location parameter $\omega_0$ and half-width at maximum $\gamma$ given by
\begin{equation}
p(\omega)=\frac{1}{\pi}\frac{\gamma}{\gamma^2+(\omega-\omega_0)^2},
\end{equation}
one can compute the order parameter $r$ explicitly 
\begin{equation}
r=
\begin{cases}
0 & \lambda<\lambda_c \\
\sqrt{1-\frac{\lambda_c}{\lambda}}, & \lambda>\lambda_c .
\end{cases}
\end{equation}
Here, $\lambda_c$ denotes a critical coupling strength given by $\lambda_c=2\gamma$.
We conclude that the system of coupled oscillations exhibits a phase-transition 
as a function of synchronization strength $\lambda$, 
characterized by a complete lack of synchronization 
below the critical coupling strength, and partial synchronization above.

\section{Physical principles of flagellar synchronization}

We now apply the generic concepts introduced above to the specific case of beating cilia and flagella, 
which can synchronize their beat by means of hydro-mechanical coupling.

The hydrodynamic flows generated by beating flagella are characterized by low Reynolds numbers \cite{Purcell:1977,Shapere:1987,Lauga:2009}.
The Reynolds number characterizes the relative importance
of inertial over viscous forces, and is defined as 
$\mathrm{Re}=\rho v L/\eta$
given a typical speed $v$ and size $L$ of active motion.
Here, $\rho$ and $\eta$ denote the density and the dynamic viscosity of the fluid. 
In the limit of zero Reynolds number, one obtains the Stokes equation, 
which for an incompressible, Newtonian fluid reads \cite{Happel:hydro}
\begin{equation} 
\label{eq_stokes}
0 = -\nabla p + \eta \nabla^2 \vbf.
\end{equation}
Here, $p$ and $\vbf$ denote pressure and flow velocity.
For the following, it will be important that the Stokes equation is linear and time-reversible, 
\textit{i.e.}\ the fluid flow $\vbf$ will simply change its sign if a flagellum would play its swimming stroke backwards in time. 
This time-reversibility has well-known consequences for self-propulsion of microswimmers at low Reynolds numbers, 
implying that a microswimmer must use a non-reciprocal swimming stroke 
that explicitly breaks time-reversal symmetry to allow for net propulsion \cite{Purcell:1977,Shapere:1987}. 
Similar arguments also apply to synchronization by hydrodynamic forces in the Stokes limit. 

\subsection{Lack of hydrodynamic synchronization in a minimal model}

\begin{figure}
\begin{center}
\resizebox{0.75\columnwidth}{!}{%
\includegraphics{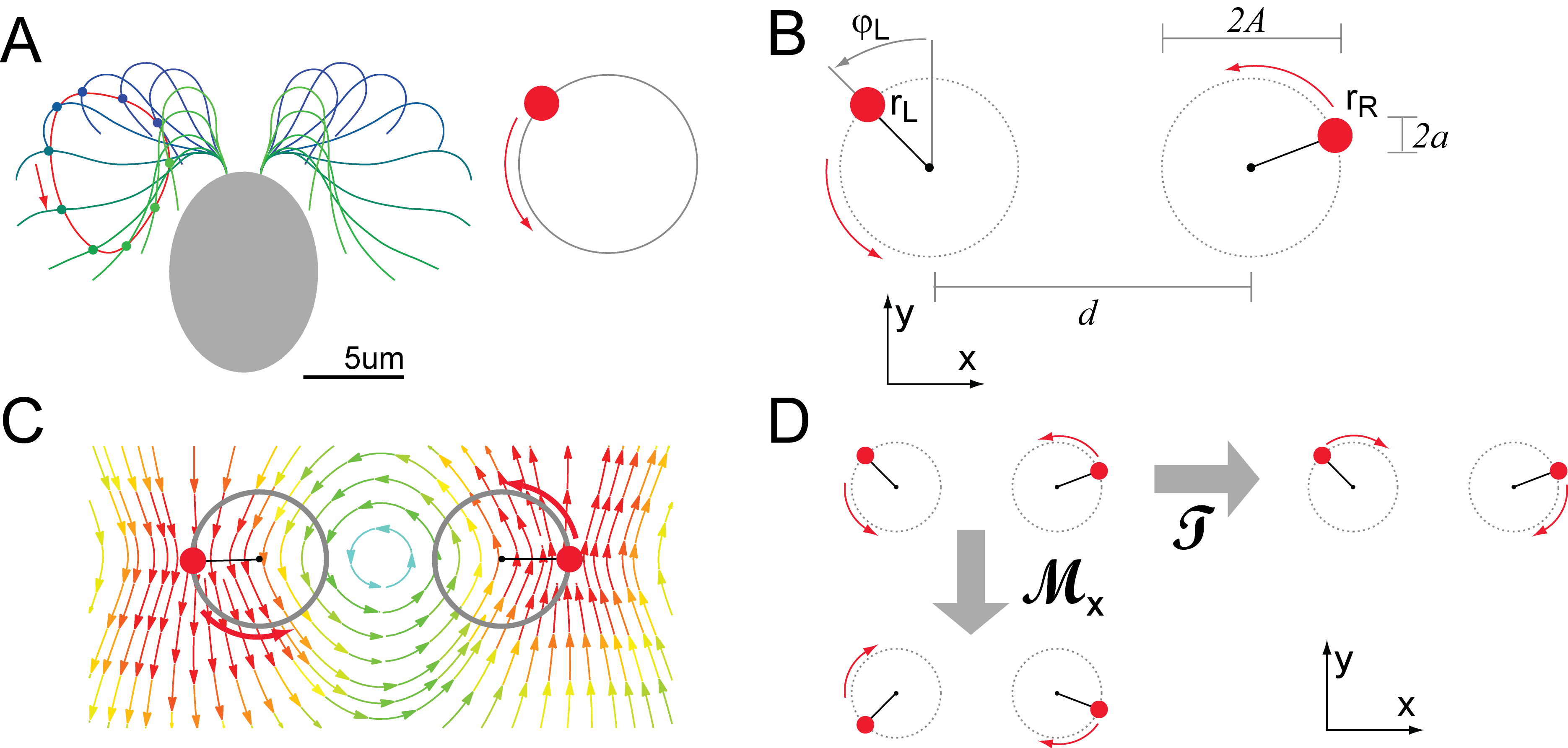}
}
\end{center}
\caption[]{
\textit{Lack of hydrodynamic synchronization in a minimal model.}
\textbf{A.}
Each point on a beating flagellum traces out a circular orbit. 
To study qualitative aspects of hydrodynamic synchronization, 
we can replace each beating flagellum by a sphere that moves along a circle.
\textbf{B.}
Two co-rotating spheres of radius $a$ move along circular tracks of radius $A$, $r_L$ and $r_R$, which are separated by a distance $d$.
Active driving forces of constant magnitude act on the spheres tangential to their track.
Additionally, constraining forces keep the spheres on track.
\textbf{C.}
The motion of the two spheres induces flow of the surrounding fluid (here indicated in color-code).
Thereby, the motion of the two spheres becomes coupled by virtue of hydrodynamic interactions:
the flow induced by the motion of one sphere propagates to the second sphere, where it exerts a hydrodynamic friction force.
\textbf{D.}
Although hydrodynamic interactions induce a weak coupling between the instantaneous motion of the two spheres, 
the net effect of this coupling is zero. 
This can be seen from symmetries: 
a reflection $\mathcal{M}_x$ at the $x$-axis, and a time-reversal $\mathcal{T}$ both change the dynamics of the spheres in the same way. 
As a time-reversal flips the stability behavior of synchronized states, while a reflection leaves this unaltered, 
we conclude that synchronization cannot occur in this minimal model, see text further explanation. 
}
\label{figure_two_spheres}
\end{figure}

Minimal models that retain key symmetries of flagellar swimming and synchronization 
have proven very useful to elucidate fundamental principles of hydrodynamic synchronization 
\cite{Taylor:1951,Vilfan:2006,Niedermayer:2008,Reichert:2005,Gueron:1999,Kim:2004,Uchida:2011}.
Specifically, a beating flagellum has been mimicked by a sphere moving along a perfect circle \cite{Vilfan:2006}, 
see Figure \ref{figure_two_spheres}A.
This description is motivated by the observation that each point on a flagellum traces out a circular orbit
with respect to the reference frame of the cell while the flagellum beats.
An entire flagellum could be described as a collection of several of such idealized spheres \cite{Geyer:2013}.

We consider the case of two spheres of radius $a$
that are constrained to move along respective circular trajectories of radius $A$, 
$\rbf_L(\varphi_L)$ and $\rbf_R(\varphi_R)$,
which are parametrized by phase angles $\varphi_L$ and $\varphi_R$ and separated by a distance $d$, 
see Figure \ref{figure_two_spheres}
\begin{equation}
\rbf_L(\varphi_L)=
A
\left( 
\begin{matrix}
\cos\varphi_L \\
\sin\varphi_L 
\end{matrix}
\right),\quad
\rbf_R(\varphi_R)=
\left(
\begin{matrix}
d \\
0
\end{matrix}
\right)
+A
\left( 
\begin{matrix}
\cos\varphi_R \\
\sin\varphi_R 
\end{matrix}
\right).
\end{equation}
Each sphere is driven by a driving force $\F_j$ of constant magnitude $f_0$ 
that acts tangential to its circular track, 
i.e.\
$\F_j = f_0 \tbf_j$,
for $j\in\{L,R\}$,
where $\tbf_j=r^{-1} \partial \rbf_j/\partial \varphi_j$ denote the local tangent vectors.
In the inertia-less limit, 
these driving forces are balanced by the hydrodynamic friction forces $\H_L$ and $\H_R$ exerted by the spheres on the surrounding fluid, 
\begin{equation}
\F_L\cdot\tbf_L = \H_L\cdot\tbf_L,\quad
\F_R\cdot\tbf_R = \H_R\cdot\tbf_R. 
\label{eq_balance}
\end{equation}
Note that force balance between driving force and hydrodynamic friction force holds only tangential to the track.
Along the normal direction, there can be a difference of forces.
In fact this force difference amounts to the constraining forces needed to keep the spheres on their circular track. 
We use short-hand notation 
$P_j=A\H_j\cdot\tbf_j$ and
$Q_j=A\F_j\cdot\tbf_j$ for  
$j\in\{L,R\}$.

\begin{figure}
\begin{center}
\resizebox{0.75\columnwidth}{!}{%
\includegraphics{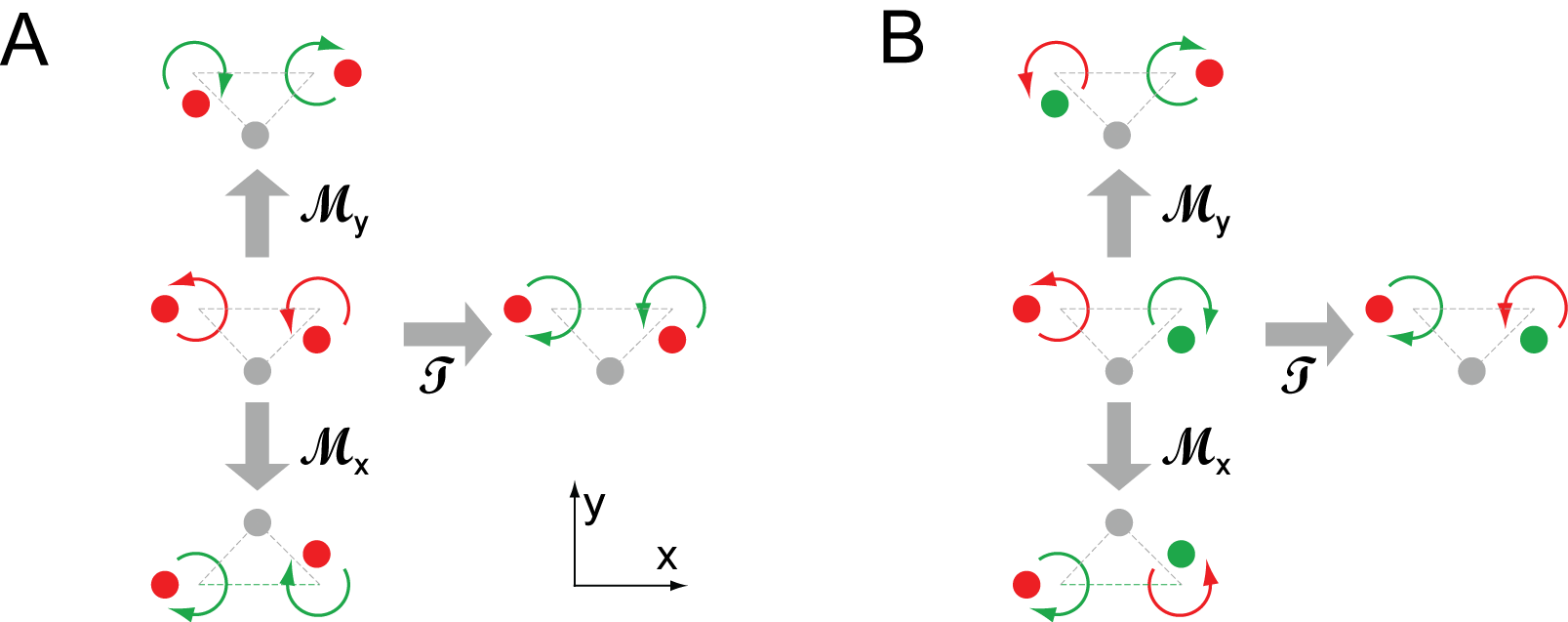}
}
\end{center}
\caption[]{
\textit{The minimal model revisited.}
The spatio-temporal symmetry of the minimal model of two co-rotating spheres 
has to be broken to allow for net synchronization by direct hydrodynamic interactions.
\textbf{A.}
A stationary third sphere (gray) breaks the dynamic equivalence between the system reflected by $\mathcal{M}_x$ 
and the system after time-reversal $\mathcal{T}$. 
However, a reflection $\mathcal{M}_y$ at the $y$-axis 
again yields a system that is dynamically equivalent to the time-reversed system. 
This second spatio-temporal symmetry rules out net synchronization for a pair of co-rotating spheres,
if the third stationary sphere is in a symmetric position as drawn here.
\textbf{B.}
The situation is different for the case of counter-rotating spheres, which more closely mimics the mirror-symmetric beat 
of the two flagella of the \textit{Chlamydomonas} cell shown in Figure \ref{figure_two_spheres}. 
In this case, either mirror operation yields a dynamics that is different from the time-reversed dynamics. 
Indeed, synchronization by hydrodynamic interactions does occur \cite{Polotzek:2013}.
}
\label{figure_three_spheres}
\end{figure}

In the limit of small spheres, $a\ll A$, and large separation distances, $d\gg A$,
the hydrodynamic friction forces are readily computed using 
the Stokes drag coefficient $\gamma=6\pi\eta a$ of a single sphere, 
and the Oseen tensor $\G(\rbf)$ that describes how the flow generated by the motion of one sphere 
propagates to the other sphere,
\begin{equation}
\H_L = \gamma \left[ \dot{\rbf}_L - \G(\rbf_R-\rbf_L) \cdot (A\dot{\varphi}_R\, \tbf_R/\gamma) \right], 
\end{equation}
and similarly for $\H_R$.
The Oseen tensor is given by $G_{ij}(\rbf) = (8\pi\eta)^{-1} (|\rbf|^{-1} + r_i r_j|\rbf|^{-3})$.
For the following, all we need is that 
the generalized friction forces $P_j$ can be written as a linear function of the phase speeds
$\dot{\varphi}_j$, \textit{e.g.}\
$P_L = \Gamma_{LL}\dot{\varphi}_L + \Gamma_{LR}\dot{\varphi}_R$.
This general form holds true without the approximation of small spheres, 
for more general geometries, and even generalizes to free-moving swimmers \cite{Polotzek:2013}.
Thus, we can write the force balance equation, Equation (\ref{eq_balance}), as
\begin{align}
\label{eq_force_balance}
\mathbf{\Gamma}(\varphi_L,\varphi_R)
\left(
\begin{matrix}
\dot{\varphi}_L \\ 
\dot{\varphi}_R 
\end{matrix}
\right)
 =
\left(
\begin{matrix}
Q_L \\
Q_R
\end{matrix}
\right).
\end{align}
This force balance equation describes two coupled oscillators and can be re-cast 
in a form similar to Equation (\ref{eq_two_oscillators}) studied above.

We can understand the synchronization dynamics solely by considering symmetries \cite{Elfring:2009}:
let $\mathcal{T}$ be the time-reversal operator, 
while $\mathcal{M}_x$ and $\mathcal{M}_y$
denote spatial mirror operations at the $x$- and $y$-axis, respectively, see Figure \ref{figure_two_spheres}.
Since there is no explicit time-dependence in the Stokes equation, Equation (\ref{eq_stokes}),
and the laws of hydrodynamics are invariant under reflections, 
the friction matrix $\mathbf{\Gamma}(\varphi_L,\varphi_R)$ must be invariant under these 
symmetry operations. 
These symmetry operations change 
the phase speeds $\dot{\varphi}_j$ and generalized driving forces $Q_j$.

% \begin{table}
% \begin{center}
% \begin{tabular}{c|cccc}
% \hline
% & $\dot{\varphi}_L$ & $\dot{\varphi}_R$ & $Q_L$ & $Q_R$ \\
% \hline\hline
% $\mathcal{T}$ & $-\dot{\varphi}_L$ & $-\dot{\varphi}_R$ & $-Q_L$ & $-Q_R$ \\ 
% $\mathcal{M}_x$ & $-\dot{\varphi}_L$ & $-\dot{\varphi}_R$ & $-Q_L$ & $-Q_R$ \\ 
% $\mathcal{M}_y$ & $-\dot{\varphi}_R$ & $-\dot{\varphi}_L$ & $-Q_R$ & $-Q_L$ \\ 
% \hline
% \end{tabular}
% \end{center}
% \caption{Effect of symmetry operations on the dynamics of two co-rotating spheres as described by Equation (\ref{eq_force_balance}).}
% \label{tab:sym}
% \end{table}

The symmetry operations $\mathcal{T}$ and $\mathcal{M}_x$ 
induce the same transformation of the dynamic equations of motion, 
Equation (\ref{eq_force_balance}). 
Initial conditions are transformed differently, though, 
$\mathcal{M}_x: \varphi_j(0)\rightarrow-\varphi_j(0)$ and 
$\mathcal{T}: \varphi_j(0)\rightarrow\varphi_j(0)$, 
yet this does not change the argument.
Thus, applying either $\mathcal{M}_x$ or $\mathcal{T}$, 
we should find the same stability behavior of synchronized steady-states. 
The crucial point is now that a time-reversal $\mathcal{T}$ 
will flip the stability behavior of any synchronized steady-state:
a stable steady-state will become stable and vice versa. 
In the Adler equation, Equation (\ref{eq_Adler}), 
this corresponds to $\lambda\rightarrow -\lambda$.
On the other hand, a spatial mirror operation will not change the stability behavior of any steady-state. 
We conclude that there cannot be any stable or unstable steady-state of the untransformed system.
In fact, all steady-states are neutrally stable, corresponding to the case $\lambda=0$
in Equation (\ref{eq_Adler}).

Thus, the symmetries of the Stokes equation rule out stable synchronization in this minimal model.
A number of specific generalizations have been proposed, all of which break the $\mathcal{T}\mathcal{M}_x$-symmetry 
in a specific manner as discussed in the next section. 

\paragraph{Note on generalized forces.}
The quantities $P_j$ above represent generalized forces with units of a torque,
which are conjugate to $\varphi_j$ in the sense of Lagrangian mechanics of dissipative systems 
\cite{Goldstein:mechanics,Vilfan:2009,Polotzek:2013}. 
Formally, these generalized friction forces are defined as
$P_j=\partial\mathcal{R}/\partial\dot{\varphi}_j$,
where $\mathcal{R}$ denotes the total rate of hydrodynamic dissipation, 
which plays the role of a Rayleigh dissipation function.
Lagrangian mechanics is particularly useful to describe mechanical systems with constraints
and applies both in conservative and non-conservative systems with quadratic dissipation function. 

\subsection{Hydrodynamic synchronization by broken symmetries}

Already Taylor suspected that flagellar synchronization is a hydrodynamic effect \cite{Taylor:1951},
yet the first experimental demonstrations were only recent \cite{Geyer:2013,Brumley:2014}.
Again, simple model systems served as a proof of principle that synchronization by hydrodynamic forces is possible,
both in theory
\cite{Vilfan:2006,Niedermayer:2008,Reichert:2005,Gueron:1999,Kim:2004,Uchida:2011}
and experiment
\cite{Kotar:2010,Bruot:2011,Bruot:2012,Leonardo:2012,Lhermerout:2012}.
These model systems demonstrated equally clearly the need for non-reversible phase dynamics and thus broken symmetries 
\cite{Elfring:2009}.
In these models, symmetries were broken either by wall effects \cite{Vilfan:2006},
by additional elastic degrees of freedom \cite{Niedermayer:2008,Reichert:2005}, 
or by evoking phase-dependent driving forces \cite{Uchida:2011,Bruot:2012}.

\paragraph{Symmetry breaking by boundary walls.}
In the original formulation of the two-sphere-model by Vilfan \textit{et al.} \cite{Vilfan:2006}, 
a no-slip boundary close to the two circular trajectories was introduced, 
which changes the friction matrix $\mathbf{\Gamma}$, 
and in particular breaks the symmetry of $\mathbf{\Gamma}$ 
with respect to the spatial mirror-operation $\mathcal{M}_x$.
Later work used a third, stationary sphere, instead of the boundary wall \cite{Friedrich:2012c,Polotzek:2013}. 
Note that in addition to the $\mathcal{T}\mathcal{M}_x$-symmetry, 
one can make a similar symmetry argument also for a reflection $\mathcal{M}_y$ at the $y$-axis.    
To break also this second symmetry, 
elliptic trajectories with a tilted orientation with respect to the boundary wall 
were considered in the work cited to provide stable synchronization \cite{Vilfan:2006}. 
Interestingly, the second symmetry argument only applies to the case of co-rotating spheres, 
but not for counter-rotating spheres with $\omega_L=-\omega_R$, 
which makes synchronization easier in that case \cite{Friedrich:2012c,Polotzek:2013}, 
see Fig.~\ref{figure_three_spheres}.

\paragraph{Symmetry breaking by amplitude compliance.}
Lenz \textit{et al.}\ introduced an elastic compliance into the model by 
considering the radii $A_L(t)$ and $A_R(t)$ of the left and right track as additional degrees of freedom. 
Specifically, the spheres were held by elastic springs with rest length $A$ and spring constant $k$, 
which can slightly change their length in response to hydrodynamic forces.
The synchronization strength was found to be inversely proportional to the stiffness of the springs, 
$\lambda\sim 1/k$. 
This result is consistent with zero synchronization strength $\lambda\rightarrow 0$ in the limit of a hard constraint 
$A=A_0$ for $k\rightarrow\infty$.
Several experimental realizations 
highlighted the role of elastic compliance for hydrodynamic synchronization
in pairs of man-made oscillators,  
including helices rotating at low Reynolds numbers in a tank of highly viscous silicone oil \cite{Qian:2009}, 
or micro-rotors driven by laser light \cite{Leonardo:2012}.
A recent study of the synchronization of beating flagella by direct hydrodynamic interactions
also suggested an important role of elastic waveform compliance of the flagellar beat for synchronization \cite{Brumley:2014}. 

\begin{figure}
\begin{center}
\resizebox{0.75\columnwidth}{!}{%
\includegraphics{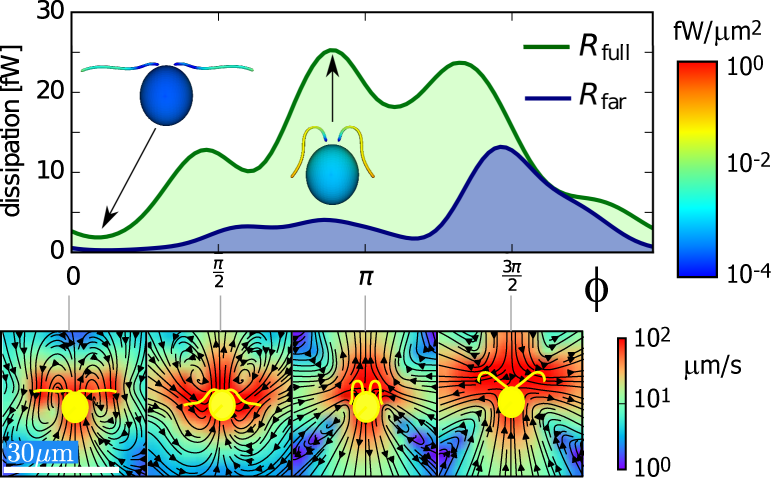}
}
\end{center}
\caption[]{
\textit{Phase-dependent hydrodynamic dissipation.}
The rate of hydrodynamic dissipation for a swimming \textit{Chlamydomonas} cell changes during the flagellar beat cycle, 
suggesting phase-dependent flagellar driving forces. 
The green curve shows the total hydrodynamic dissipation, while the blue curve depicts 
a suitably defined dissipation in the far field \cite{Klindt:2015}.
The insets in the lower row show the flagellar shapes at the different phase of the beat cycle 
together with a numeric computation of the flow profile.
The total hydrodynamic dissipation rate is maximal during the recovery stroke, when the two flagella move close to the cell body. 
In contrast a hydrodynamic dissipation rate computed for the far-field of fluid flow is maximal during the effective stroke.
In a simple description of the flagellar beat, the hydrodynamic dissipation rate
corresponds to an active driving force of the flagellar beat \cite{Geyer:2013}.
Phase-dependent active driving forces represent a general mechanism for hydrodynamic synchronization 
that explicitly breaks symmetries that rule out synchronization \cite{Uchida:2011}. 
Modified from \cite{Klindt:2015}.
}
\label{figure_chlamy_dissipation}
\end{figure}

\paragraph{Symmetry breaking by phase-dependent driving forces.}
Golestanian \textit{et al.} discussed phase-dependent driving forces $Q_j(\varphi_j)$ 
as a generic mechanism for hydrodynamic synchronization.
Such phase-dependence represent another way to break $\mathcal{T}\mathcal{M}_x$-symmetry.
In computational models of flagellar synchronization for the bi-flagellated green alga \textit{Chlamydomonas}, 
phase-dependent driving forces were indeed found to contribute to synchronization \cite{Bennett:2013,Geyer:2013}.
In short, synchronization arises from a ``resonance'' between a phase-dependence of the active driving 
and the phase-dependence of the hydrodynamic friction coefficients.
In a coarse-grained theoretical description of flagellar beating, 
phase-dependent driving forces of the flagellar beat can be computed from the phase-dependent rate of hydrodynamic dissipation \cite{Geyer:2013,Klindt:2015}.
Figure \ref{figure_chlamy_dissipation} shows the phase-dependence of the hydrodynamic dissipation rate computed 
for a free-swimming \textit{Chlamydomonas} cell.

\paragraph{Symmetry breaking by inertial effects.}
Theers \textit{et al.} considered the effect of small, but non-zero Reynolds number $\mathrm{Re}$ \cite{Theers:2013}. 
The presence of unsteady acceleration terms in the Navier-Stokes equation breaks it 
time-reversal symmetry and thus allows for stable synchronization. 
The synchronization strength scales as $\lambda\sim (\mathrm{Re})^{1/2}$.
A similar line of argument should also apply to complex fluids 
with visco-elastic properties or nonlinear constitutive relations, 
which likewise break time-reversal symmetry. 

\subsection{Flagellar synchronization independent of hydrodynamic interactions}

\begin{figure}
\begin{center}
\resizebox{0.95\columnwidth}{!}{%
\includegraphics{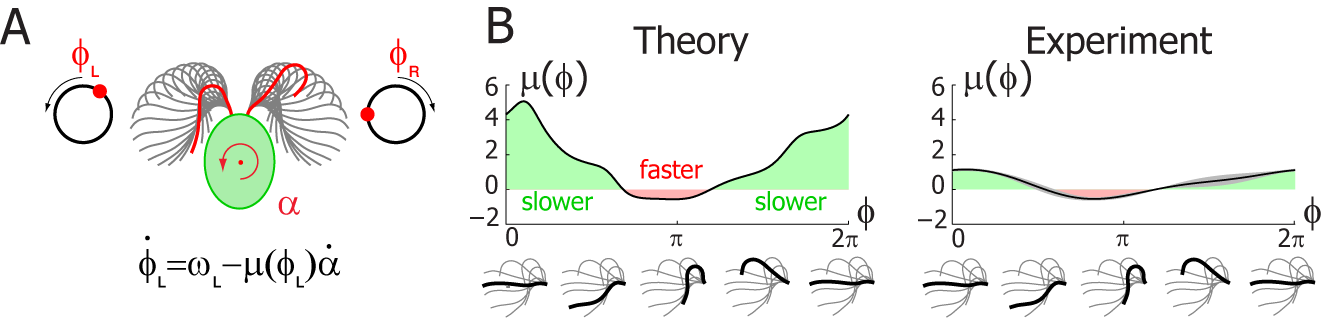}
}
\end{center}
\caption[]{
\textit{Dynamic force-velocity-relation of the flagellar beat}
\textbf{A.}
In a theoretical description of flagellar swimming and synchronization in the green alga \textit{Chlamydomonas}, 
each beating flagellum was described by a phase oscillator with respective phases $\varphi_L$ and $\varphi_R$ \cite{Geyer:2013}.
In this description, the flagellar phase speed $\dot{\varphi}_j$ changes in response to a change in hydrodynamic load, 
which can in particular result from a rotation of the entire cell with rotation rate $\dot{\alpha}$.
In the absence of any rotation, $\dot{\alpha}=0$, the flagellum beats at its intrinsic frequency $\omega_L$. 
For a rotating cell, additional hydrodynamic friction forces act on the flagellum to which
the flagellum responds by a change in phase speed with susceptibility $\mu(\varphi)$.
This susceptibility characterizes a dynamic force-velocity relationship of the beating flagellum.
\textbf{B.}
The coupling coefficient $\mu(\varphi)$ can be computed in terms of hydrodynamic friction forces, 
assuming that the flagellar oscillator operates at full load with negligible internal dissipation. 
The susceptibility has been measured in experiments with \textit{Chlamydomonas} cells
that displayed oscillatory rotational motion due to a lack of flagellar synchronization.
During the effective stroke of the left flagellum ($\varphi\approx 0$), 
a counter-clockwise rotation of the entire cell ($\dot{\alpha}>0$)
increases the motion of the flagellum relative to the fluid at rest. 
As a consequence, the hydrodynamic friction forces acting on this flagellum increase, and the flagellum slows down ($\mu>0$), 
both in theory and experiment. 
During the recovery stroke, the effect is opposite: 
The active shape change of the flagellum and a passive counter-clockwise rotation partially cancel each other.
This reduces the hydrodynamic friction forces acting on the left flagellum, causing it to speed its beat up ($\mu<0$).
% Note that the theory used in \cite{Geyer:2013} has been parametrized from a data set of synchronized flagellar beating, where the cell did not rotate, 
% and is thus free from adjustable parameters.
% One may extend this previous theory by introducing a single fit parameter, which describes dissipation inside the flagellum;
% this additional fit parameter allows to adjust the amplitude of the theoretically predicted coupling coefficient $\mu(\varphi)$ 
% to the amplitude of the experimentally measured coupling (unpublished). 
Modified from \cite{Geyer:2013}.
}
\label{fig_force_velocity}
\end{figure}

In addition to mechanisms for flagellar synchronization 
that rely on direct hydrodynamic interactions between two flagella
as discussed in the previous section, 
Friedrich \textit{et al.} recently introduced an alternative mechanism
that is independent of hydrodynamic interactions,
but relies instead on a coupling between flagellar beating and swimming motion \cite{Friedrich:2012c,Geyer:2013}.
It was proposed that this synchronization mechanism applies in particular to free-swimming \textit{Chlamydomonas} cells.
In short, both the left and right flagellum of a \textit{Chlamydomonas} cell were
modeled as phase-oscillators with respective phase, $\varphi_L$ and $\varphi_R$. 
The speed of the beat $\dot{\varphi}_j$, $j\in\{L,R\}$, 
depends on the instantaneous hydrodynamic load acting on the flagella,
which in turn depends on the swimming motion of the cell, 
characterized by its rate of rotation $\dot{\alpha}$
\begin{align}
\label{eq_phi_chlamy}
\dot{\varphi}_L &= \omega_0 - \mu(\varphi_L) \dot{\alpha}, \\
\dot{\varphi}_R &= \omega_0 + \mu(\varphi_R) \dot{\alpha}.
\end{align}
These dynamical equations can be derived from simple assumptions using 
a general framework of force balances.
In particular, the coupling function $\mu$ can be computed for any given beat pattern
in terms of hydrodynamic forces acting on the two flagella, see Figure~\ref{fig_force_velocity}.
Direct hydrodynamic interactions between the two flagella 
can be accounted for, but have been neglected in Equation (\ref{eq_phi_chlamy}) for sake of simplicity.
Asynchronous beating of the two flagella causes a rotation of the cell
with rotation rate $\dot{\alpha}$ 
\begin{equation}
\label{eq_alpha}
\rho(\varphi_L,\varphi_R)\dot{\alpha} = \nu(\varphi_L) \dot{\varphi}_L -\nu(\varphi)_R\dot{\varphi}_R.
\end{equation}
Equation (\ref{eq_alpha}) represents a torque balance equation 
that balances the torques for a rigid body rotation of the entire cell (on the left-hand side)
and the sum of the torques generated by the beat of the left and right flagellum, respectively (right-hand side).
Note that Equation \ref{eq_alpha} implies $\dot{\alpha}=0$ if both flagella beat in-phase, 
as expected by symmetry.

We can re-arrange the dynamical system given by Equations (\ref{eq_phi_chlamy}) and (\ref{eq_alpha})
into the form of an Adler equation for the phase difference $\delta=\varphi_L-\varphi_R$, see Equation (\ref{eq_Adler}).
This allows to express the effective synchronization strength $\lambda$ in terms of the various coupling functions \cite{Geyer:2013}
\begin{equation}
\lambda=-\oint_0^{2\pi} d\varphi\, \frac{2\mu(\varphi)\nu''(\varphi)}{\rho(\varphi,\varphi)-2\mu(\varphi)\nu(\varphi)}
\end{equation}
The coupling functions $\mu$, $\nu$, and $\rho$ can in turn be computed in terms of hydrodynamic friction forces. 
For wild-type beat patterns, it was calculated $\lambda=0.3$, which implies stable in-phase synchronization \cite{Geyer:2013}.
Note that alternative beat patterns can result in $\lambda<0$, which renders the in-phase synchronized state unstable, 
and favors instead anti-phase synchronization with $\delta\approx\pi$.

\subsection{The beating flagellum is a noisy oscillator}

The flagellar beat represents a biological oscillator with small, yet perceivable fluctuations
\cite{Polin:2009,Goldstein:2009,Goldstein:2011,Ma:2014,Wan:2013,Wan:2014,Werner:2014}.
By mapping the periodic shape deformations of the flagellar beat on the normal form of a Hopf oscillator,
Equation (\ref{eq_hopf_noise}), it is possible to quantify both phase and amplitude fluctuations \cite{Ma:2014}.
In the simplified case of a noisy phase oscillator, given by Equation (\ref{eq_phi_noise}), 
we can compute the quality factor of noisy oscillations 
\begin{equation}
Q=\frac{\omega_0}{2D_\varphi}.
\end{equation}
The quality factor is a measure after how many oscillation cycles
the oscillator phase decorrelates due to noise.
For bull sperm flagella, 
it was found 
$Q=38.0\pm 16.7\,(\mathrm{mean}\pm\mathrm{s.e.})$ \cite{Ma:2014}.
This value is on the same order-of-magnitude as previous, 
indirect measurements for \textit{Chlamydomonas} flagella
based on the frequency of phase-slips in pairs of synchronized flagella \cite{Goldstein:2009}. 
The corresponding noise strength $D_\varphi=\omega_0/(2Q)$  is several orders of magnitude larger 
than an estimate for contribution from passive, thermal fluctuations 
based on a fluctuation-dissipation theorem \cite{Ma:2014}. 
Thus, fluctuations of the flagellar beat are of active origin \cite{Goldstein:2009,Ma:2014}.
Fast fluctuations of the flagellar beat with correlation times on a millisecond time-scale 
have been attributed to small-number-fluctuations in the collective dynamics of the approximately $10^5$ molecular motors 
inside the sperm flagellum that drive its beat \cite{Ma:2014}.
Thus, fluctuations of the flagellar beat represent mesoscopic signatures of stochastic activity of individual molecular motors.
Subsequent, more refined analyses of flagellar fluctuations have shown that amplitude fluctuations are phase-dependent,
being minimal during the initiation of bends at the proximal end of the flagellum \cite{Wan:2014,Werner:2014}.
Additionally, slow fluctuations of the flagellar beat with correlation times on a second time-scale have been observed, 
which may reflect noise in chemical signaling cascade that control the shape of the flagellar beat \cite{Wan:2014}.

The synchronization of several flagella represents a stochastic phenomenon 
that arises from a competition between a mechanical coupling (which favors phase-locking), 
and the effects of noise (which counter-act synchronization) \cite{Polin:2009,Goldstein:2009,Goldstein:2011,Ma:2014}.  

\section{Discussion}
Flagellar synchronization represents a model system for the collective dynamics of biological oscillators.
Experiments have demonstrated that a mechanical coupling by viscous forces can indeed synchronize several flagella \cite{Brumley:2014}.
Theory has outlined different physical mechanisms for hydrodynamic synchronization \cite{Vilfan:2006,Niedermayer:2008,Uchida:2011,Theers:2013},
each of which breaks the time-reversal symmetry of hydrodynamics at low Reynolds numbers in a different way.
Which of the different mechanisms dominates in different biological systems will have to determined by future research. 
Open questions relate in particular to the role of elastic anchorage of the flagellar apparatus \cite{Quaranta:2015},
and the waveform compliance of the flagellar beat \cite{Wan:2014}.
It is not known, which features of the flagellar beat patterns 
determine whether in-phase or anti-phase synchronization will be stable, 
as observed \textit{e.g.} in flagellar mutants \cite{Leptos:2013}. 

Previous theory suggested that states of synchronized dynamics correspond to either a maximum or a minimum of the 
rate of hydrodynamic dissipation \cite{Elfring:2009}. 
However, it is not known if this rule is universally applicable and which features determine
which extremum corresponds to a stable synchronized state.

We are only beginning to understand how the flagellar beat responds to time-varying external forces \cite{Geyer:2013,Wan:2014}.
Such an understanding will be important not only to understand synchronization in collections of beating flagella by mechanical coupling, 
but may also be informative on the microscopic mechanisms of motor control that regulate flagellar bending waves
\cite{Lindemann:1994,Brokaw:2008,Riedel:2007}.
Future research can bridge the gap between educative minimal models of flagellar synchronization,
and the complexity of biological systems.

%% bibliography
% \def\urlprefix{}
% \def\url#1{}
% \bibliography{../../bibliography/library}

% \end{document}

\end{document}